\documentstyle[twoside,fleqn,espcrc2,epsf]{article}

\newcommand{\nn}{\nonumber}
\def\lsim{\raise0.3ex\hbox{$<$\kern-0.75em\raise-1.1ex\hbox{$\sim$}}}
\def\gsim{\raise0.3ex\hbox{$>$\kern-0.75em\raise-1.1ex\hbox{$\sim$}}}

\newcommand{\pslash}{p\kern-1ex /}
\newcommand{\Dslash}{{\cal D}\kern-1.5ex /}

\newcommand{\vev}[1]{\left\langle #1 \right\rangle}

\newcommand{\J}[4]{{#1} {\bf #2} (#3) #4}

\newcommand{\NP}{Nucl.~Phys.}

\newcommand{\PL}{Phys.~Lett.}
\newcommand{\PR}{Phys.~Rev.}

\title{Eigenvalues of the hermitian Wilson-Dirac operator
and chiral properties of the domain-wall fermion%
\thanks{Talk presented by K.-I.\ Nagai}}

\author{CP-PACS Collaboration :
  A.~Ali~Khan\rlap,\address{Center for Computational Physics,
    University of Tsukuba, Tsukuba, Ibaraki 305-8577, Japan}
  S.~Aoki\rlap,\address{Institute of Physics,
    University of Tsukuba, Tsukuba, Ibaraki 305-8571, Japan}
  Y.~Aoki\rlap,$^{\rm a,b}$\thanks{address after 1 May, 2000:
        RIKEN BNL Research Center, Brookhaven National
        Laboratory, Upton, NY 11973, USA}
  R.~Burkhalter\rlap,$^{\rm a,b}$
  S.~Ejiri\rlap,$^{\rm a}$
  M.~Fukugita\rlap,\address{Institute for Cosmic Ray Research,
    University of Tokyo, Kashiwa 277-8582, Japan}
  S.~Hashimoto\rlap,\address{High Energy Accelerator Research Organization
    (KEK), Tsukuba, Ibaraki 305-0801, Japan}
  N.~Ishizuka\rlap,$^{\rm a,b}$
  Y.~Iwasaki\rlap,$^{\rm a,b}$
  T.~Izubuchi\rlap,\address{Institute of Theoretical Physics, Kanazawa
    University, Ishikawa 920-1192, Japan}
  K.~Kanaya\rlap,$^{\rm b}$
  T.~Kaneko\rlap,$^{\rm d}$
  Y.~Kuramashi\rlap,$^{\rm d}$
  T.~Manke\rlap,$^{\rm a}$\thanks{address after 1 Feb., 2000:
        Department of Physics, Columbia University,
        538 W 120th St., New York, NY 10027, USA}
  K.-I.~Nagai\rlap,$^{\rm a}$
  J.~Noaki\rlap,$^{\rm a}$
  M.~Okawa\rlap,$^{\rm d}$
  H.P.~Shanahan\rlap,$^{\rm a}$\thanks{address after 15 Sept., 2000:
        Department of Biochemistry and Molecular
        Biology, University College London, London, England, UK}
  Y.~Taniguchi\rlap,$^{\rm b}$
  A.~Ukawa$^{\rm a,b}$ and
  T.~Yoshi\'e$^{\rm a,b}$
}

\begin{document}

\begin{abstract}
Chiral properties of QCD formulated with the domain-wall fermion 
(DWQCD) are studied using the anomalous quark mass $m_{5q}$ 
and the spectrum of the 4-dimensional Wilson-Dirac operator.  
Numerical simulations are made with the standard plaquette gauge 
action and a renormalization-group improved gauge action. 
Results are reported on the density of zero eigenvalue obtained with the 
accumulation method, and a comparison is made with the results for $m_{5q}$.
\end{abstract}

\maketitle

\section{Introduction}
\label{sec:intro}
Formulation of chiral fermions on the lattice 
has been one of long-standing problems 
in lattice field theories.
Several years ago, 
the domain-wall fermion (DWF) formalism~\cite{Kaplan,Shamir},
which is a Wilson fermion in $D+1$ dimensions 
with Dirichlet boundary condition in the extra dimension,  
has been proposed as a new formulation of lattice chiral fermion.
In the limit of large extra dimension size, $N_s\rightarrow\infty$, 
the spectrum of free domain-wall fermion
contains massless modes at the edges in the extra dimension. 

While the massless modes are shown to be stable 
in perturbation theory\cite{AokiTanig,KikuNeuYama},
their existence may be spoiled non-perturbatively in the presence of
dynamical gauge fields.
We studied this issue through an anomalous quark mass $m_{5q}$
in Ref.~\cite{cppacs}.   
This quantity measures the magnitude of chiral symmetry breaking with 
the domain-wall QCD (DWQCD). 

In this article we make a status report of our attempt to understand 
the results on the $N_s$-dependence of $m_{5q}$ obtained in 
Ref.~\cite{cppacs} through measurements of the eigenvalue distribution of 
the 4-dimensional Wilson-Dirac operator. 

\section{Chiral Properties of DWQCD}
\label{sec:m5q}

We define the anomalous quark mass by\cite{cppacs}
\begin{equation}
m_{5q} = \lim_{t\to\infty}
\frac{\sum_{\bf x}\left<J_{5q}^a(t,{\bf x})P^b(0,{\bf 0})\right>}
        {\sum_{\bf x}\left<P^a(t,{\bf x})P^b(0,{\bf 0})\right>}.
\label{eq:m5q}
\end{equation}
This quantity measures the chiral symmetry breaking effect in 
the axial Ward-Takahashi identity:
\begin{eqnarray}
\sum_\mu \vev {\nabla_\mu A_\mu^a(x) P^b(0)} = 
2m_f \vev{P^a(x) P^b(0)} \nn \\
+2\vev{J_{5q}^a(x) P^b(0)}
+i\vev{\delta^a_x P^b(0)} ~,
\label{eqn:WTid}
\end{eqnarray}
where $A_\mu^a(x)$ is the axial-vector current,
$P^a(x)$ is the pseudoscalar density, 
and $J_{5q}^a(x)$ represents the explicit breaking of chiral symmetry.
On a smooth gauge field background, 
the anomalous contribution $\vev{J_{5q}^a(x) P^b(y)}$ 
vanishes as $\exp (-c N_s)$ at large $N_s$ \cite{FurmanShamir}. 
Therefore $m_{5q}$ also vanishes exponentially in this case. 

\begin{figure}[t]
\centerline{\epsfxsize=7.5cm \epsfbox{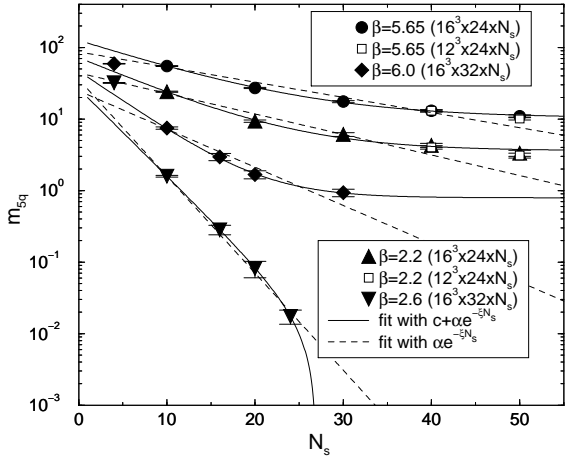}}
\vspace{-1cm}
\caption{$m_{5q}$ as a function of $N_s$ 
at $a^{-1} \simeq 1$ GeV and $m_0=1.7$
and at $a^{-1} \simeq 2$ GeV and $m_0=1.8$, where $m_0$ 
is the domain-wall height, for the plaquette ($\beta=5.65, 6.0$) 
and RG-improved ($\beta=2.2, 2.6$) gauge action \protect\cite{cppacs}.
}
\label{fig:m5q}
\vspace{-3mm}
\end{figure}

In Ref.~\cite{cppacs}, we carried out quenched simulations 
to study the $N_s$ dependence in detail.  
Two values of lattice spacing, $a^{-1} \simeq  1$ and 2~GeV, 
are explored, using both the plaquette and an RG-improved gauge actions.

Our main results for $m_{5q}$ are summarized in Fig.~\ref{fig:m5q}.
The results for the plaquette action are obtained at $\beta=5.65$ 
($a^{-1}\simeq 1$~GeV) and 6.0 (2~GeV), 
and those for the RG-improved action at $\beta=2.2$ (1~GeV) and 
2.6 (2~GeV).  Solid lines are fits to $c+\alpha e^{-\xi N_s}$, and 
dashed lines to $\alpha e^{-\xi N_s}$ 

Important points to note in Fig.~\ref{fig:m5q} are:
(i) comparing results for the spatial sizes $12^3$ and $16^3$ at 
$a^{-1}\simeq 1$~GeV, we find 
the finite volume effects in $m_{5q}$ to be small, 
(ii) $m_{5q}$ decreases with the lattice spacing, 
(iii) the magnitude of $m_{5q}$ is smaller for the RG-improved action 
than for the plaquette action, and 
(iv) from data in the range of $N_s$ we explore,
$m_{5q}$ seems to remain non-zero
in the limit $N_s \rightarrow \infty$, 
in all cases except at $\beta=2.6$ for the RG-improved action.
If confirmed with studies at larger values of $N_s$, the last point means
that DWQCD realizes chiral symmetry
at $a^{-1}\simeq 2$~GeV only for the case of the RG-improved action.

\section{Eigenvalues of the hermitian Wilson-Dirac operator 
and chiral property}
\label{sec:eigenvalue}

Chiral symmetry of DWQCD can be studied also 
through the transfer matrix in the direction of the extra 
dimension\cite{Neuberger,KikuNogu}. 
When the transfer matrix has a unit eigenvalue, 
chiral symmetry is not realized in DWQCD 
because the left and right chiral modes on the two 
edges in the extra dimension couple with each other. 

A unit eigenvalue of the transfer matrix is in one-to-one correspondence 
with a zero eigenvalue of the hermitian Wilson-Dirac operator 
defined by 
\begin{eqnarray}
H_W(m_0) = \gamma_5 D_W(-m_0) ~, 
\end{eqnarray}
which is much easier to calculate.
Here, $D_W(-m_0)$ is the four dimensional Wilson-Dirac kernel 
with a bare mass $-m_0$. 
Therefore, a failure of exponential decay of $m_{5q}$ would result 
if $H_W$ develops a zero eigenvalue.

We calculate eigenvalues of $H_W^2$ by the Lanczos method using 
50--100 configurations at several values of coupling in the range
$a^{-1} \simeq 1$--2 GeV 
using both plaquette and RG-improved actions.  
The results from the Lanczos method are checked by the Ritz functional 
method for $H_W$.
We also study the dependence on the lattice size. 
The maximum lattice at $a^{-1} \simeq 1$~GeV is $12^4$ for both actions,
while the one at $a^{-1}\simeq 2$~GeV is $24^4$ for the RG-improved action
and $16^3 \times 32$ for the plaquette action.

\subsection{Eigenvalue distributions}

\begin{figure}[t]
\centerline{\epsfxsize=7.5cm \epsfbox{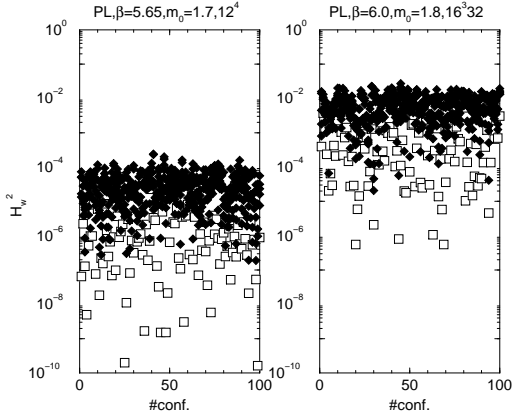}}
\vspace{-1cm}
\caption{
Monte Carlo time histories for the six lowest eigenvalues of 
$H_W^2$ obtained with the plaquette gauge action. 
}
\label{fig:evdistP}
\vspace{-3mm}
\end{figure}

\begin{figure}[t]
\centerline{\epsfxsize=7.5cm \epsfbox{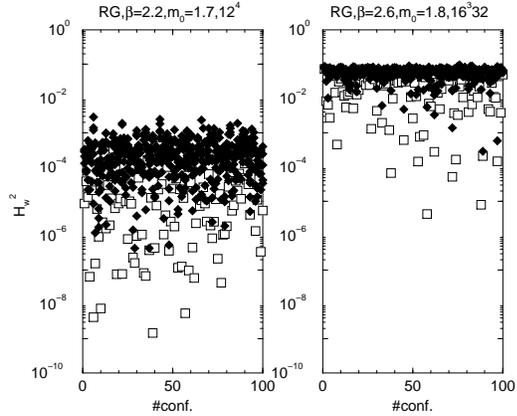}}
\vspace{-1cm}
\caption{
The same as Fig.~\protect\ref{fig:evdistP} obtained with the RG-improved 
gauge action. 
}
\label{fig:evdistR}
\vspace{-3mm}
\end{figure}

In Figs.~\ref{fig:evdistP} and \ref{fig:evdistR}
we plot Monte Carlo time histories for the six lowest eigenvalues of 
$H_W^2$ for the plaquette and RG-improved actions.  
In each figure the left panel shows results for $a^{-1}\simeq 1$~GeV 
and the right panel for $a^{-1}\simeq 2$~GeV.  
The lattice size at $a^{-1} \simeq 2$~GeV is the same 
as in the previous work of $m_{5q}$ shown in Fig.~\ref{fig:m5q}.
Open squares plot the minimum eigenvalue $\lambda^2_{\rm min}$ and 
filled diamonds are the five higher eigenvalues.

There is a clear trend that the minimum eigenvalues become larger 
for smaller lattice spacings.  Another interesting point is that 
the RG-improved action gives larger values of $\lambda^2_{\rm min}$ than the 
plaquette action, which indicates that the RG-improved action has a 
better chiral behavior.  These trends are parallel to the features 
we noted for $m_{5q}$ in Sec.~\ref{sec:m5q}.

\subsection{Spectral density}

The spectral density of $H_W$ is defined by 
\begin{equation}
\rho(\lambda) = \lim_{V \rightarrow \infty}
\frac{1}{3\cdot 4\cdot V}\sum_{\lambda'}\delta (\lambda'-\lambda),
\end{equation}
where the summation is over the eigenvalues of $H_W$. 
We are interested in the density of zero eigenvalues, $\rho(0)$, 
since we expect this quantity to be related to the existence of unit 
eigenvalue of the transfer matrix. 
To calculate this quantity, 
we adopt the accumulation method proposed in \cite{EHN}, 
which is based on the relation
\begin{eqnarray}
A(\lambda) & \equiv & \int_0^{\lambda^2} d \lambda^{{\prime}^2}
\widetilde{\rho}(\lambda^{{\prime}^2})
= \frac{1}{3 \cdot 4 \cdot V} 
\sum_{|\lambda^\prime| \le \lambda} 
\bf{1}\\
& = & \int_{-\lambda}^\lambda d \lambda^\prime \rho(\lambda^\prime) 
\simeq 2 \rho(0) \lambda  + O(\lambda^2),
\label{eq:ansatz}
\end{eqnarray}
where $\widetilde{\rho}(\lambda^2)$ is the spectral density function
for $H_W^2$. 
We note that, for the small-$\lambda$ expansion of $A(\lambda)$ 
in (\ref{eq:ansatz}), 
analyticity of $\rho(\lambda)$ at the origin is assumed.

\begin{figure}[t]
\centerline{\epsfxsize=7.5cm \epsfbox{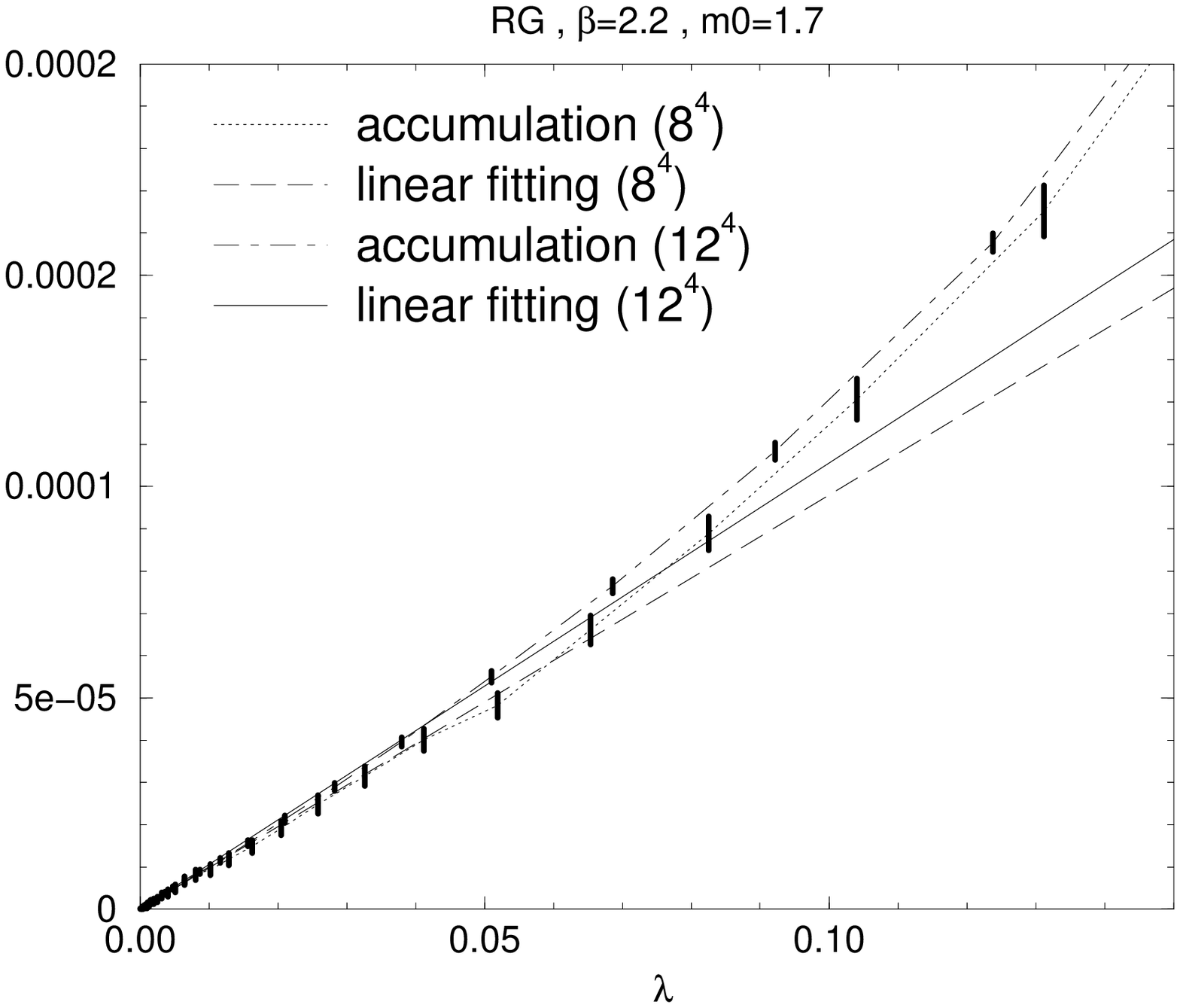}}
\vspace{-0.3cm}
\centerline{\epsfxsize=7.5cm \epsfbox{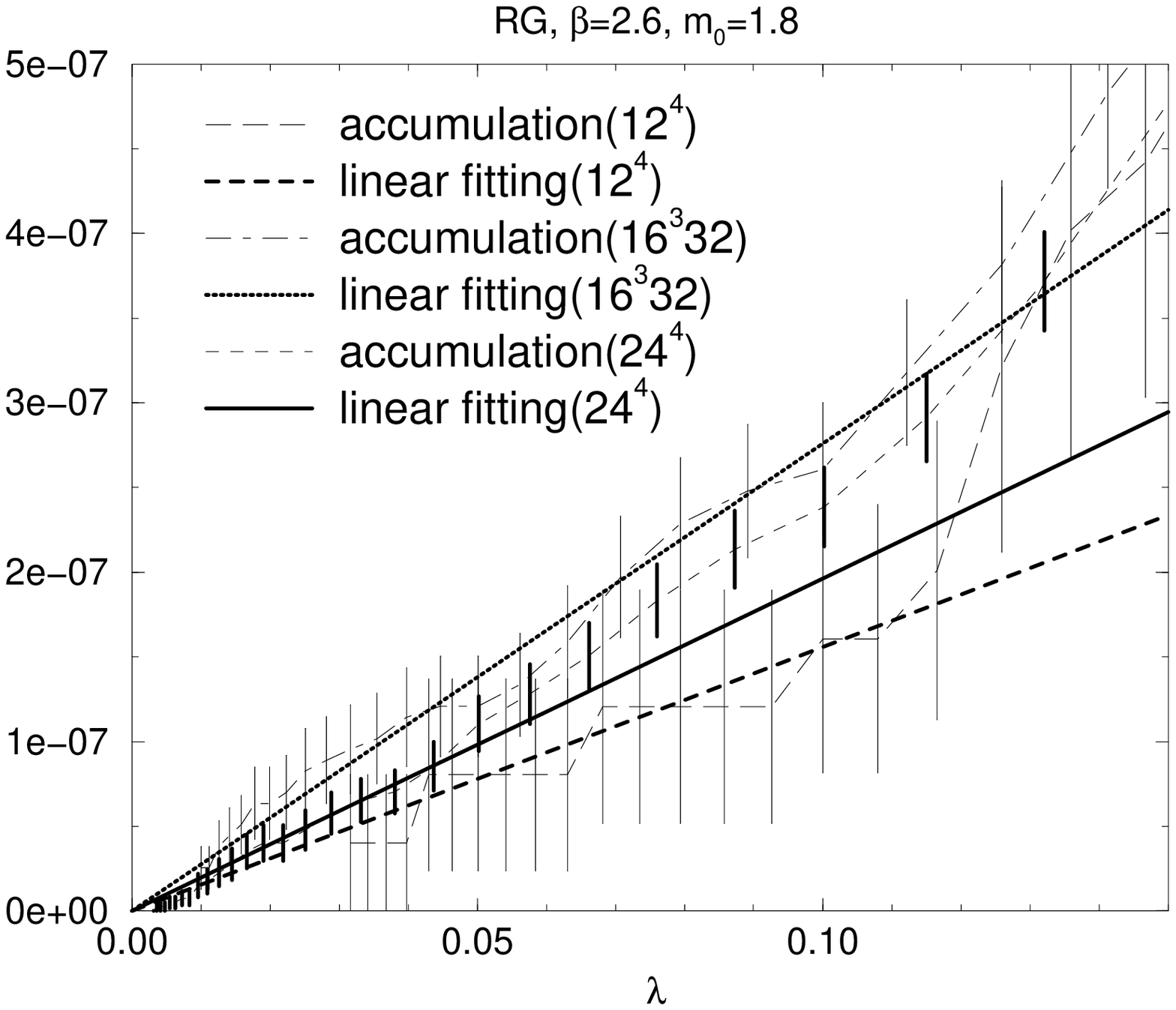}}
\vspace{-1.2cm}
\caption{
The accumulation $A(\lambda)$ at $\beta=2.2$ 
($a^{-1}\simeq1$ GeV) and 
$\beta=2.6$ ($a^{-1}\simeq2$ GeV) from the RG-improved action.
}
\label{fig:acc}
\vspace{-3mm}
\end{figure}

\begin{figure}[t]
\centerline{\epsfxsize=7.5cm \epsfbox{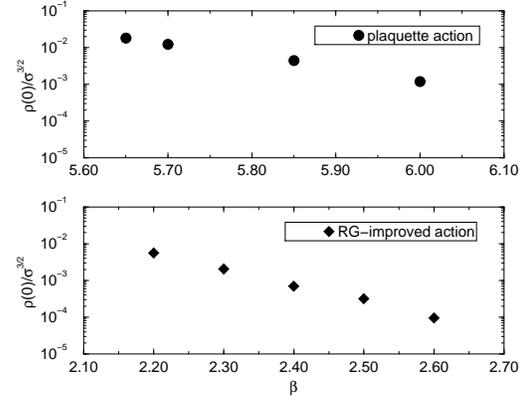}}
\vspace{-1cm}
\caption{
$\rho(0)/\sigma^{3/2}$ as a function of $\beta$
for the plaquette and RG-improved gauge action. 
Data obtained on the largest lattices are shown.
}
\label{fig:rhob}
\vspace{-3mm}
\end{figure}

In Fig.~\ref{fig:acc}, we show typical examples of the 
accumulation $A(\lambda)$ from the eigenvalue distribution of $H_W^2$
for the case of the RG-improved action.
Results for $\rho(0)$ obtained by a linear fitting following 
(\ref{eq:ansatz}), normalized by the string tension, are 
summarized in Fig.~\ref{fig:rhob}. 

Our results for $\rho(0)$ for the plaquette action are consistent with 
the previous data by Edwards {\it et al.} \cite{EHN}.
Results for the RG-improved action show a similar $\beta$ dependence. 
A significant difference is that the RG-improved action leads to much 
smaller values of $\rho(0)$ than the plaquette action, roughly by 
an order of magnitude. 

\section{Discussions}
\label{sec:discussions}

We have applied the accumulation method to estimate 
the spectral density at zero eigenvalue
of the hermitian Wilson-Dirac operator, $\rho(0)$.
We found that this method leads to non-zero values of $\rho(0)$ at 
$a^{-1}\simeq1$--2 GeV for both the plaquette and RG-improved actions. 

At $a^{-1}\simeq1$ GeV, the non-zero result for $\rho(0)$ is  
consistent with the finite $m_{5q}$ in the large $N_s$ limit observed 
in \cite{cppacs} with both the plaquette and RG-improved actions.  
At $a^{-1}\simeq2$ GeV, while a consistency also holds 
with the plaquette action, 
there is an apparent contradiction for the case of the RG-improved action 
since $m_{5q}$ seems to decay exponentially with $N_s$ for this case. 

In the accumulation data shown in Fig.~\ref{fig:acc}
we observe that results are very noisy at $\beta=2.6$ (2~GeV).  
Since the fit results for $\rho(0)$ fluctuates with volume, 
it is difficult to determine the size dependence.
Therefore simulations with larger lattices are needed to 
check if the slope remains non-vanishing toward infinite volume. 
Another point to examine is if the analyticity assumption for 
$\rho(\lambda)$ at the origin is justified if there is a 
spectral gap.  Further studies are required to clarify these points. 

\vspace{2mm}
This work is supported 
in part~by~Grants-in-Aid~of~the~Ministry~of~Education~(Nos.\
10640246,~10640248,~10740107,~11640250, 
11640294,~11740162,~12014202,~12304011, 12640253,~12740133).
AAK and TM are supported by JSPS Research for the Future Program
(No.\ JSPS-RFTF 97P01102).
SE, TK, KN, JN and HPS are JSPS Research Fellows.

%

%
%

\end{document}